\documentclass[numberedappendix]{emulateapj}
\usepackage{psfig}

\newcommand{\fig}[1]{Figure~\ref{#1}}
\newcommand{\tab}[1]{Table~\ref{#1}}

\newcommand{\sex}{\texttt{SExtractor}}
\newcommand{\igalfit}{\texttt{iGalFit}}
\newcommand{\isex}{\texttt{iSEx}}
\newcommand{\galfit}{\texttt{GalFit}}
\newcommand{\dsnine}{\texttt{ds9}}
\newcommand{\sersic}{S\'ersic}
\shorttitle{\igalfit\ User's Manual}
\shortauthors{Ryan Jr.}

\begin{document}

\title{\igalfit: An Interactive Tool for GalFit}

\author{R. E. Ryan Jr.\altaffilmark{*,**}}

\email{rryan@stsci.edu}

\altaffiltext{*}{Physics Department, University of California, Davis, CA 95616}
\altaffiltext{**}{Space Telescope Science Institute, Baltimore, MD 21218}

\begin{abstract}

We   present  a   suite   of  IDL   routines   to  interactively   run
\galfit\ whereby  the various  surface brightness profiles  (and their
associated parameters)  are represented by regions, which  the User is
expected to  place.  The regions may  be saved and/or  loaded from the
ASCII  format used  by \dsnine\  or  in the  Hierarchical Data  Format
(version 5).  The  software has been tested to run stably  on Mac OS X
and  Linux with  IDL~7.0.4.  In  addition  to its  primary purpose  of
modeling  galaxy  images  with   \galfit,  this  package  has  several
ancillary uses,  including a flexible image  display routines, several
basic    photometry    functions,    and    qualitatively    assessing
\texttt{SExtractor}.  We distribute the package freely and without any
implicit or explicit warranties,  guarantees, or assurance of any kind.
We kindly ask  users to report any bugs, errors,  or suggestions to us
directly  (as opposed  to fixing  them themselves)  to  ensure version
control and uniformity.

\end{abstract}

\keywords{methods: data analysis --- techniques: image processing 
  --- galaxies: structure --- galaxies: fundamental parameters}

\section{Introduction} \label{intro}

The  shape,  size,  and  structure  of distant  galaxies  can  provide
invaluable  insight  to their  formation  history. Consequently,  many
there  have been  many techniques  and codes  developed to  make these
measurements:               \texttt{GIM2D}              \citep{mar98},
\galfit\  \citep{peng02,peng10},  \texttt{GASPHOT} \citep{pig06},  and
\texttt{GALPHAT} \citep{yoon11}.  While this is  in no way meant to be
an exhaustive list, it merely highlights the interest in, and emphasis
placed on,  robustly measuring  the properties of  the two-dimensional
light distributions of galaxies.

As  astronomical surveys have  grown ever  wider, samples  of galaxies
have become larger, and accordingly these detailed modeling techniques
have also  evolved.  Turning  toward a ``pipeline  approach,'' whereby
many sophisticated  programs are called  in concert to  streamline the
measurements               on               large              samples
\citep[e.g.~\texttt{GALAPAGOS},][]{hau11},   many   authors  sacrifice
detailed fitting for bulk properties.  Naturally, such a paradigm will
invariably generate  a series of tunable  parameters to be  set by the
User,  many   of  which,  can  significantly  alter   the  success  or
reliability of  the pipeline.   While appropriate settings  are likely
obvious or easily ascertained, they are often tailored to a particular
sample, and as the sample  changes so must the settings.  For example,
Users often  need to provide the  shape modeling codes  a source list,
since rarely  do these codes  also identify objects.   Obviously, this
places  the  utmost  importance  on  the  identification  scheme,  for
multiple reasons: First and most  obviously, if the object of interest
is failed to  be cataloged, then clearly the power  of the pipeline is
for  not.    Secondly,  if   the  identification  software   fails  to
``deblend'' a  neighboring object,  which presumably should  be either
simultaneously modeled  or masked from  the fitting, then  the results
are not to be trusted.  Both  of these issues (and many others) can be
mitigated by human-intervention or supervision throughout the process,
which is the primary motivation for this work.

In  this  article, we  present  our  software,  \igalfit, which  is  a
graphical user  interface (GUI) for running \galfit.   The software is
inspired     by     the     successful     image     display     tool,
\texttt{ds9}\footnote{http://hea-www.harvard.edu/RD/ds9/}   (and   its
predecessors), produced  and maintained  by the Chandra  X-ray Science
Center.   In  \igalfit, the  Users  is  expected  to place  (circular,
elliptical,  and rectangular)  regions on  the image  to  indicate the
function to be fit and the  initial guesses of the parameters. In this
way, the  User has complete  control over the  critical identification
step, while abdicating the  assembly-line power of a pipeline.

This article should be treated  as somewhat of a ``User's Manual'' and
a  reference point  for  the  package.  The  article  is organized  as
follows: in Section~\ref{igalfit} we  describe the basics of the code,
in Section~\ref{sextractor}  we briefly  describe a second  package to
integrate  \sex\ with \galfit,  in Section~\ref{ancillary}  we discuss
potential  ancillary  uses,  and  in Section~\ref{future}  we  mention
several upgrades for future versions.

\section{Interactive \galfit: \igalfit} \label{igalfit}
\subsection{Installing \igalfit}\label{install}

The   entire   package   is    written   in   the   Interactive   Data
Language\footnote{http://www.ittvis.com/language/en-us/productsservices/idl.aspx}
(hereafter  IDL),  therefore  having   IDL  installed  is  an  obvious
prerequisite.   With  IDL installed,  the  installation is  relatively
straightforward:
\begin{enumerate}
  \item    Obtain    the    IDL    routines   for    \igalfit\    from
    http://dls.physics.ucdavis.edu/$\sim$rer/   or  by   emailing  the
    author.
  \item  Create an environment  variable in  your start-up  file named
    \texttt{igalfit}, and  set it  equal to the  full-path of  the IDL
    routines.
  \item Amend the \texttt{IDL\_PATH} variable to include this newly set
    variable.  Obviously, the variable should be set {\bf ABOVE} the 
    \texttt{IDL\_PATH} setting.
  \item For the Mac OS X platform, configure the ``Apple Key'' for the
    keyboard accelerators used by \igalfit:
    \begin{enumerate}
      \item In the  User's home directory, there may  be a file called
        \texttt{.Xdmodmap} (if not, create one) and add the following 
        lines:
        \begin{itemize}
          \item clear mod1
          \item clear mod2
          \item add mod1 $=$ Meta\_$\,$L
        \end{itemize}
        \item Start the X11 server and open the {\it Preferences} tab.  
          Under the {\it Input} dialog, make the the following items 
          are {\bf UNCHECKED}:
          \begin{itemize}
            \item Follow system keyboard layout
            \item Enable key equivalents under X11
          \end{itemize}
    \end{enumerate}
    \item  To  install  the   optional  Perl  script  which  can  call
      \igalfit\ from the command line,  simply include the path to the
      Perl executable in your start-up file.
    \item Restart the X11 server.
\end{enumerate}

\begin{figure*}
\centerline{\psfig{file=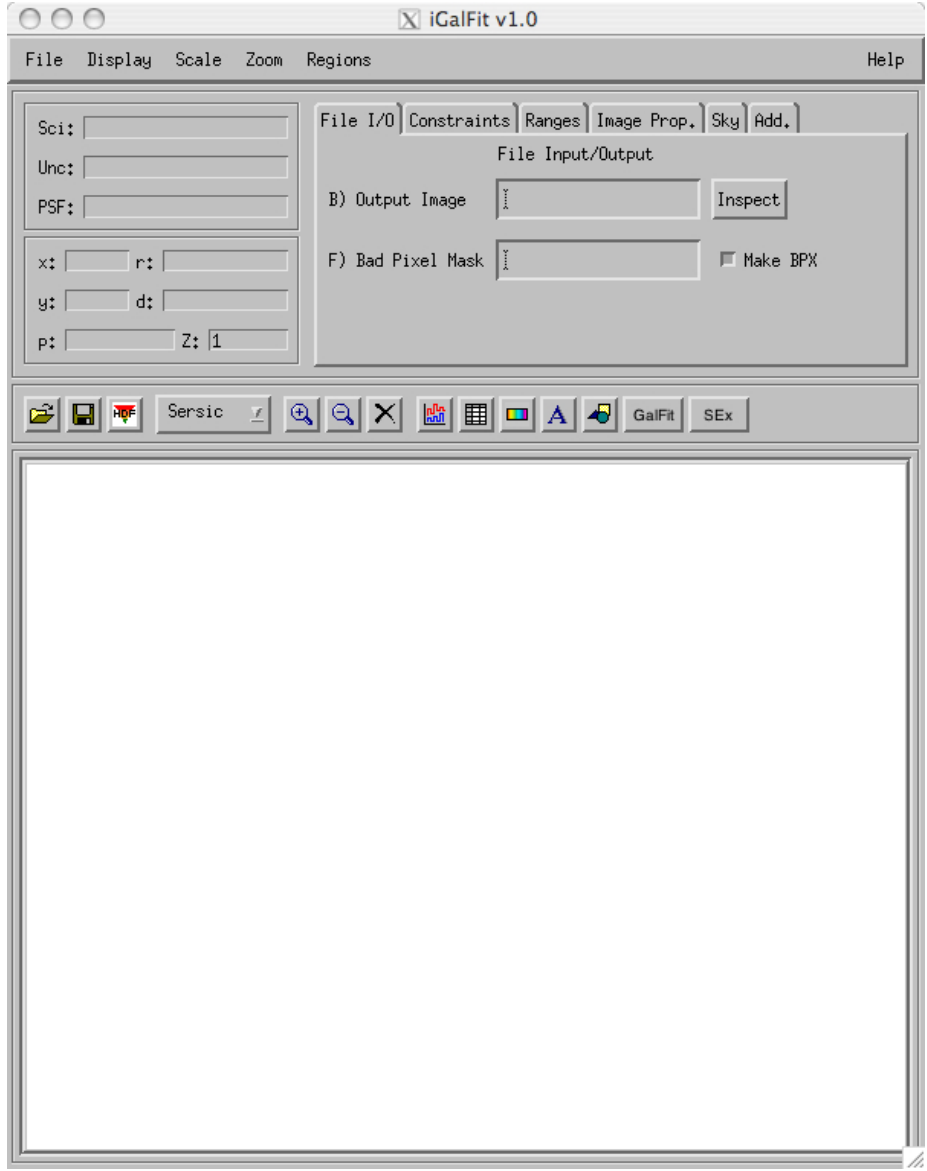,width=5.0in}}
\caption{Screenshot of \igalfit.  We show \igalfit\ in its initial
  state: no image, regions, or sub-GUIs present. As discussed in
  \S~\ref{start}, many of the fields can be set programmatically from
  the IDL prompt or the command-line via our Perl script. \label{igf}}
\end{figure*}

\subsection{Starting \igalfit} \label{start}
Since the code is primarily written in IDL, the simplest way to start 
\igalfit is  to issue the command ``igalfit'' at the IDL prompt:
\begin{verbatim}
IDL> igalfit, [OPTIONS=options]
\end{verbatim}
This of course requires that IDL is running, which can be a nuisance.
Therefore we  include a brief  Perl script, which will  initialize IDL
and run  \igalfit\ and can be  run from the command-line  in the usual
way.  There are several optional  keywords, which can be set in either
the IDL or Perl, to (pre-)set various items. A complete listing can be
given  by typing ``igalfit,/help''  in IDL,  ``igalfit -help''  at the
command line, or are given in \tab{options}.

\begin{center}
\begin{table}
\caption{\igalfit\ Optional Inputs}
\label{options}
\begin{tabular*}{0.5\textwidth}
  {@{\extracolsep{\fill}}llp{2.5in}}
\hline\hline
\multicolumn{1}{l}{IDL$^{\dagger}$} & \multicolumn{1}{l}{Perl} & \multicolumn{1}{l}{Function}\\

\hline
LOADSETTINGS & -load & Load an \igalfit\ save file.\\
SCIFILE & -sci & Load the science image.\\
UNCFILE & -unc & Load the uncertainty image.\\
PSFFILE & -psf & Load the PSF image.\\
BPXFILE & -bpx & Load the bad-pixel image.\\
IMGFILE & -img & Load the output image.\\
MAGZERO & -zero & Set the magnitude zeropoint.\\
\hline
\multicolumn{3}{p{3.4in}}{$^{\dagger}$While IDL is not case-sensitive, we follow the 
convention that optional keywords are in all-caps.}\\
\end{tabular*}
\end{table}
\end{center}

\subsection{Controls}\label{controls}

The controls to \igalfit\ are  largely modeled after those of \dsnine,
and  so users  should be  able  to seamlessly  move between  programs.
However, we will describe the control system for completeness:

\begin{description}
\item[File Input/Output:] While many  of the individual files needed to
  run  \galfit\ can be  saved/loaded at  any time,  users may  find it
  convenient to save/load  a file which encodes the  complete state of
  \igalfit.   This  facilitates  an  easy recall  or  programmatically
  assigning  a state  of \igalfit.   These  files are  written in  the
  Hierarchical Data  Format version 5 (HDF5), though  by default named
  {\it igalfit\_save.h5}.  These files  can be saved/loaded by buttons
  on the toolbar or the {\it  File} pulldown menu, or with the command
  line options (see \tab{options}).
\item[Region  Input/Output:] The  regions for  \igalfit\  indicate the
  positions, sizes, and morphologies of the objects to be fit, regions
  to  be masked,  and  allowable fitting  regions  (discussed in  more
  detail in \S~\ref{rungf}.  Users can save/load regions from the {\it
    Regions} pulldown  menu, and are  meant to be  directly compatible
  with \dsnine.
\item[Mouse Functions:] The mouse controls many aspects of \igalfit:
  \begin{description}
    \item[Left click  and drag:]  If on a  blank region of  the image,
      then a new  region based on the current  state of the morphology
      pulldown menu in  the primary toolbar (see \fig{igf}).   If on a
      valid region, then this will select the region allowing the user
      to  translate  or  delete  the  region and  initiate  the  region
      ``handles.''   The handles can  be selected  to adjust  the size
      (left  click) and  rotate  (left click  while  holding the  {\sc
        shift} key).
    \item[Middle click:] Recenter on the current mouse position.
    \item[Right click  and drag:] Adjust  the ``stretch'' of  the color
      map ---  vertical and horizontal  movements adjust the  bias and
      contrast, respectively.
    \item[Left  double-click:] If  on a  region, then  the  {\it Region
      Information} sub-GUI will open.
  \end{description}
\item[Keyboard Commands:] When  the cursor is in the  main display window
  there are several actions which  can be run by keyboard actions.  We
  briefly   describe   the    valid   functions   in   \tab{keyboard}.
  \begin{center}
\begin{table}
\caption{\igalfit\ Keyboard Commands}
\label{keyboard}
\begin{tabular*}{0.5\textwidth}
  {@{\extracolsep{\fill}}lp{3.0in}}
\hline\hline
\multicolumn{1}{l}{Key} & \multicolumn{1}{l}{Function}\\

\hline

$+$ & Zoom in on the center of the display.\\
$-$ & Zoom out on the center of the display.\\
q & Close \igalfit.\\
{\sc Delete} & Delete the selected region.\\
r & Display a radial profile of object closest to cursor position.\\
m & Display image statistics for a small area around the cursor position.\\
l & Display a line plot for some region around the cursor position.\\
c & Display a column plot for some region around the cursor position.\\
h & Display a pixel histogram for some region around the cursor position.\\
e & Display a contour plot for some region around the cursor position.\\
\hline
\end{tabular*}
\end{table}
\end{center}

\item[Toolbar Functions:] Immediately above  the main display window is
  a row of buttons which  control several commonly used functions (see
  \fig{igf}).   We briefly  describe the  function of  each  button in
  \tab{icons}.

\begin{center}
\begin{table}
\caption{\igalfit\ Toolbar}
\label{icons}
\begin{tabular*}{0.5\textwidth}
  {@{\extracolsep{\fill}}lp{3.0in}}
\hline\hline
\multicolumn{1}{l}{Icon} & \multicolumn{1}{l}{Function}\\

\hline\\[-0.8em]

\includegraphics[height=1em]{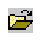} & Load an \igalfit\ save file.\\
\includegraphics[height=1em]{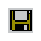} & Save an \igalfit\ save file.\\
\includegraphics[height=1em]{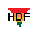}  & Inspect an \igalfit\ save file.\\

\includegraphics[height=1em]{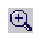}  & Zoom in on center of window.\\
\includegraphics[height=1em]{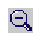}  & Zoom out on center of window.\\
\includegraphics[height=1em]{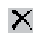}  & Delete currently selected region.\\

\includegraphics[height=1em]{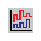} & View pixel histogram and set pixel 
min/max for display.\\
\includegraphics[height=1em]{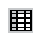} & View a pixel values under cursor (the 
size of the table can be set in the {\it Preferences} menu.\\
\includegraphics[height=1em]{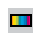} & Manually adjust the display 
settings (minimum, maximum, bias, and contrast).\\
\includegraphics[height=1em]{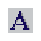} & Manually edit the \galfit\ input 
file.\\
\includegraphics[height=1em]{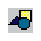} & Inspect and modify the properties 
of the loaded regions.\\

\includegraphics[height=1em]{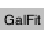} & Run \galfit\ with the current 
state.\\

\includegraphics[height=1em]{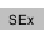} & Launch the \sex\ sub-GUI.\\

\hline
\end{tabular*}
\end{table}
\end{center}

\end{description}

\subsection{Modeling Galaxy Profiles}\label{rungf}

The primary  motivation for developing \igalfit\  was to interactively
create the input files  for \galfit\ (e.g.~bad-pixel masks, constraint
files,  ``{\it  galfit.feedme}'',  and  uncertainty  maps).   In  this
section, we describe  the typical order-of-operations to interactively
model galaxy profiles.
\begin{enumerate}
\item In principle, \igalfit\  is capable of displaying large (roughly
  10~k$\times$10~k pixels).   However, manipulating a  large number of
  pixels is computationally expensive and is generally unnecessary (or
  ill-advised) for running \galfit.  Therefore, we recommend operating
  on images sized for each object (roughly 1~k$\times$1~k pixels).

  While \galfit\ is  capable of running without a  PSF and uncertainty
  map, these images  are essential for robust estimates  of the galaxy
  profiles.  Furthermore, \galfit\ is  more reliable when operating on
  images which  are in  the units of  counts (cts;  Peng priv.~comm.),
  despite the more  common convention to process images  in count rate
  (cts/s).   Therefore the  user should  set the  appropriate exposure
  time and  unit in  the {\it Image  Properties} tab.  Finally,  for a
  meaningful estimate of the $\chi^2$ and parameter uncertainties, the
  uncertainty  image  should  represent  all sources  of  uncertainty.
  Often  times,  particularly  with  data  from  \texttt{MultiDrizzle}
  \citep{koek02}, the weight/uncertainty maps  do not include any shot
  noise from the  objects.  If this is the case,  then the user should
  set  this  flag and  \igalfit\  will  modify  the uncertainty  image
  (${\cal U}_{i,j}$) according to
\begin{equation}
{\cal U}_{i,j}\rightarrow\sqrt{{\cal U}_{i,j}^2+\left|{\cal S}_{i,j}\right|},
\end{equation}
where ${\cal S}_{i,j}$  is the science image ---  both the science and
uncertainty images have units of counts.

\item After  loading a science  image (and PSF and  uncertainty maps),
  the  user  is  expected  to  indicate  the  initial  conditions  for
  \galfit\ by  drawing regions, which represent the  model that should
  be fit.   In \tab{regions}, we list the  currently supported regions
  and their functions.   There are three ways a  user may draw regions
  in \igalfit:

\begin{center}
\begin{table}
\caption{\igalfit\ Regions}
\label{regions}
\begin{tabular*}{0.5\textwidth}
  {@{\extracolsep{\fill}}llcl}
\hline\hline
\multicolumn{1}{c}{Shape} & \multicolumn{1}{c}{Color} & 
\multicolumn{1}{c}{Rotate} & \multicolumn{1}{c}{Purpose}\\
\hline
ellipse   & green    & Y & Fit \sersic\ function\\
          & blue     & Y & Fit ExpDisk function\\
          & red      & Y & Fit DeVauc function\\
          & skyblue  & Y & Fit Nuker function\\
          & cyan     & Y & Fit Edge-on Disk function\\
          & seagreen & Y & Fit King function\\
          & orange   & Y & Fit Gaussian function\\
          & maroon   & Y & Fit Moffat function\\
          & black    & Y & Mask section\\         
circle    & magenta  & N & Fit emprical PSF\\
          & black    & N & Mask section\\         
rectangle & yellow   & N & Define a fitting section\\
          & black    & Y & Mask section\\          
\hline
\end{tabular*}
\end{table}
\end{center}

\begin{description}
\item[Manual:] The  primary advantage of  \igalfit\ is the  ability to
  interactively place the model profiles. In this way, the fitting can
  be done iteratively by modifying  properties of the regions, such as
  the color (see  \tab{regions}), position, adding/removing additional
  regions, or masking objects.

\item[Automated:]  Despite the  obvious advantages  with interactively
  placing regions,  this can  be very tedious  and daunting  for large
  images with  many objects.  Therefore,  we include a  second package
  which will call \sex\ to  identify objects and draw the appropriate
  regions  (discussed  in  more  detail in  \S~\ref{sextractor}).   To
  determine which  fitting function to assume  (see \tab{regions}), we
  have a crude star/galaxy separation algorithm of:
\begin{eqnarray}\label{autopsf}
\left.\begin{array}{ll}
    \texttt{ISOAREA\_\,IMAGE}\leq A_{\rm crit}\\
    R\geq R_{\rm crit}
\end{array}
\right\}\mathrm{PSF}
\end{eqnarray}
\begin{eqnarray}\label{automask}
\left.\begin{array}{ll}
    \texttt{ISOAREA\_\,IMAGE}\leq A_{\rm crit}\\
    R\leq R_{\rm crit}
\end{array}
\right\}\mathrm{Mask}
\end{eqnarray}
where   we  define   $R\!=\!\sqrt{\pi   ab}$  and   $(a,b)$  are   the
semi-(major,minor) axes, respectively.   We consider objects which are
considerably  smaller in area  and extent  than the  PSF to  be likely
image defects or unrejected cosmic  rays, and in which case, should be
masked  in  the  \galfit\  calculations.  All  remaining  objects  are
considered to be extended, and are assigned a single fitting function.
The  tunable parameters  $(A_{\rm  crit}, R_{\rm  crit})$ and  default
extended source function  can be set in the  {\it Preferences} menu in
\igalfit.   We caution that  these classifications  are only  meant to
guide the user in placing the  regions, and should not be trusted as a
robust morphological indicator.

\item[File:] Regions  can be loaded  from a standard  \dsnine\ regions
  file.  In Appendix~\ref{regfile}, we give an example regions file to
  illustrate the format.

\end{description}
\item With  the fitting regions  indicating the initial  conditions of
  \galfit\ set, the user should consider setting two ancillary regions.
  First,  any  pixel  which   is  seriously  corrupted  by  non-object
  flux\footnote{The  sky pixels  will be  modeled by  the \textit{sky}
    function, and  therefore should not  be masked.}  (such  as cosmic
  rays, image  defects, diffraction  spikes, bleeds, etc.)   should be
  masked.  If  bad pixels are left  unmasked, then they  will bias the
  estimate  of the  sky brightness  by \galfit,  which  will adversely
  affect other  parameters (notably the radius)  and/or give incorrect
  results for the objects directly (such as position, total magnitude,
  or radius).  Second, users should consider setting a \textit{Fitting
    Section} region,  which restricts the \galfit\  calculation to the
  interior  pixels.   If  the  fitting   section  is  not   set,  then
  \igalfit\ will fit the entire image.
\item  Once the  fitting regions  have  been placed,  the user  should
  consider  defining  any  possible  constraints.  While  the  use  of
  constraints  are   discouraged,  they  do  have   some  utility  ---
  particularly  when fitting  composite functions  (such as  bulge and
  disk).  Obviously,  any parameter which  is found by \galfit\  to be
  fixed on a  constraint border is dubious, and  the user should relax
  the constraint.  In  most cases, if a constraint  can be avoided, it
  is safest to do so.
\item  At  any  point  in  the  process,  the  user  can  inspect  the
  \galfit\ input  file (e.g.~{\it galfit.feedme}) and  make any manual
  modifications.  We caution, there  are no safe-guards to verify that
  these files do not contain any errors.
\item  If  \galfit\  successfully  runs  to  completion,  the  default
  behavior is to load  the output file (e.g.~{\it imgblock.fits}) into
  a  sub-GUI  to  display  the results  (see  \fig{inspectgf}).   This
  sub-GUI will display the science, model, and residual images (on the
  same   stretch),  the  best-fit   results,  and   global  properties
  (e.g.~degrees  of  freedom, $\chi^2$,  sky  properties, etc.).   The
  various output images can also be displayed in the main GUI.
\end{enumerate}

\section{Source Extractor} \label{sextractor}

Since  \galfit\ does  not identify  objects, the  user is  required to
indicate approximate  positions of all  sources, whether to be  fit or
masked.   In many deep  galaxy images,  the number  of objects  may be
overwhelming to  mark every source.  Therefore, we  include a separate
GUI     to    run    \texttt{Source     Extractor}    \citep[hereafter
  \sex;][]{bert96}.  While  this package  can be run  independently of
\igalfit,  the main  features  are only  accessible through  \igalfit.
There  are a  multitude  of parameters  which  govern the  operations,
however we  restrict control to only  the few which  are most commonly
modified.

The \sex\ GUI  employs two sub-GUIs to set  the measurements (i.e. the
{\it *.param}  file) and  build a convolution  filter (i.e.~  the {\it
  *.conv}  files). The  parameters  needed for  \igalfit\  are set  by
default.  The  convolution routine  allows for several  common filters
(Gaussian, Mexican  hat, top  hat, delta-function) and  a user-defined
function  of  a single  parameter.   As  of  the time  preparing  this
document, \sex\ (version 2.5.0  2009-09-30) did not permit any filters
larger than  32~pixels, however \isex\ will  not warn the  use of this
potential issue.

\begin{figure}
\psfig{file=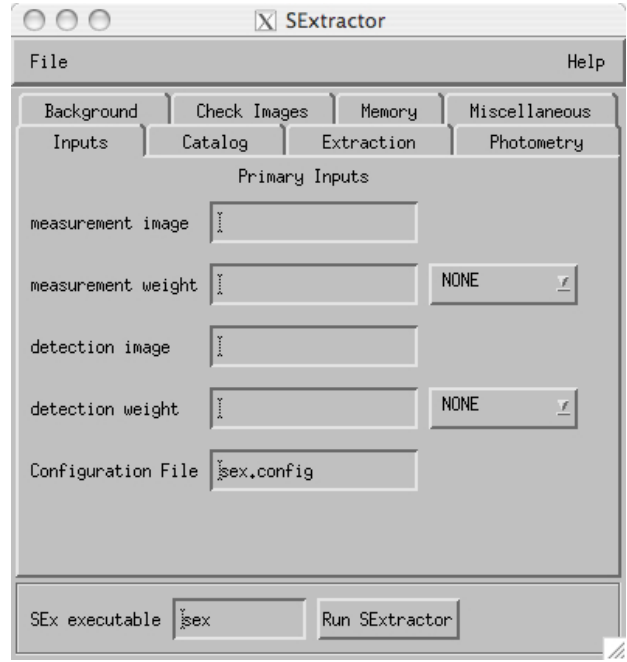,width=3.4in}
\caption{A screenshot  of the \sex\  GUI.  Each stanza of  the default
  configuration  file is  represented as  a separate  tab.  Additional
  sub-GUIs are included to create convolution filters (i.e.~the *.conv
  files)  and  select  parameters  to be  measured  (i.e.~the  *.param
  files).  While this GUI can  be run independently of \igalfit, it is
  most useful when  used in conjunction with \igalfit.   The \sex\ GUI
  initializes  with   the  parameters  necessary   to  interface  with
  \igalfit,   and  the   user   is  discouraged   from  changing   the
  \texttt{CATALOG\_TYPE}   or  removing  any   measurement  parameters
  (though adding parameters is acceptable).}
\end{figure}

\section{Ancillary Uses}\label{ancillary}
While the primary purpose of \igalfit\ is to model the two-dimensional 
light distributions of various objects, we suggest other applications 
which users may find valuable.

\subsection{Interactively Assessing \sex\ Settings}\label{runsex}

\sex\  has  become the  {\it  de  facto}  standard for  detecting  and
measuring a  number of properties  of faint objects,  particularly for
deep-field  surveys.  Not surprisingly,  there are  a host  of tunable
parameters  to  be  set  by  the  user  which  govern  the  detection,
deblending,  measurement, memory usage,  and outputting.   Therefore a
typical session  involving \sex\ begins with  a trial-and-error period
of  tweaking various  parameters until  the output  catalog satisfies
some qualitative property, often times the detection/deblending of the
faintest sources.  Despite the  various manuals ({\it Source Extractor
  for   Dummies}:  B.~Holwerda,   {\it   SExtractor  User's   Manual}:
E.~Bertin), there  can be a great deal  of confusion on the  role of a
given  parameter and  how  it  can be  affected  by other  parameters,
particularly for  novice users.  Since \igalfit\  can directly control
\sex\ using  a separate interface,  users are able to  experiment with
any combination of \sex\ settings and their effects.

\subsection{Inspect Images} \label{inspect}

Since the controls and image display aspects of \igalfit\ was inspired
by \dsnine,  it has  a many flexible  quick-look tools  built-in.  The
images  can be  scaled  and stretched,  zoomed,  panned, and  separate
frames  (although  this would  be  implemented  by loading  additional
images  into the  PSF and  Uncertainty  fields) for  easy display  and
comparisons.  At  present, rotations are not supported,  but this will
likely  be   included  in   subsequent  versions.   As   mentioned  in
\S~\ref{rungf},  the  display and  processing  of  large images  (~10k
$\times$10k pixels) is ill-advised.

\begin{figure*}
\centerline{\psfig{file=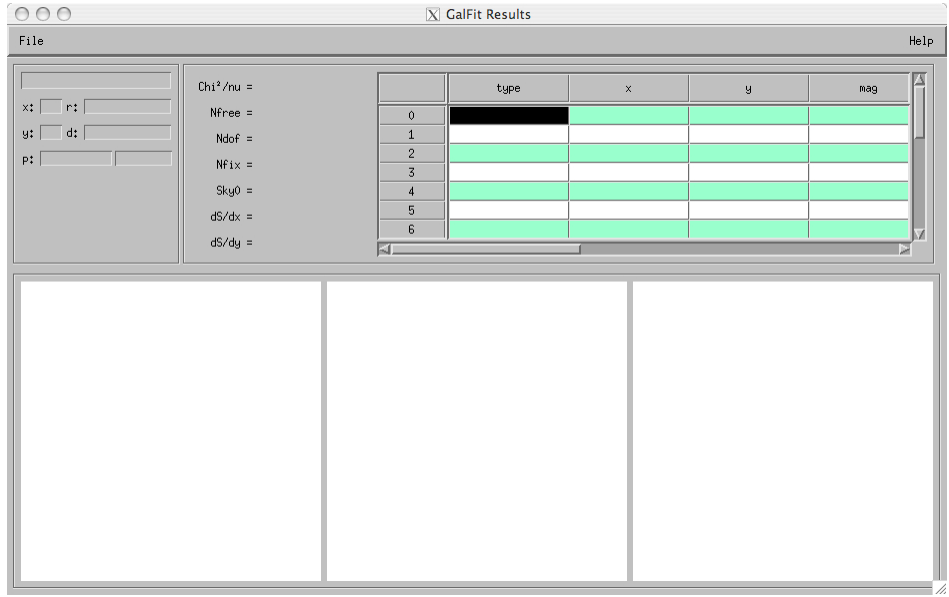,width=5.2in}}
\caption{Screenshot of  the GUI to inspect the  \galfit\ results.  All
  of the information displayed here  is taken from the \galfit\ output
  file (i.e.~the {\it imgblock.fits}).  This  GUI has many of the same
  controls and mouse functionality as \igalfit.\label{inspectgf}}
\end{figure*}

\subsection{Quick-Look Photometry} \label{photo}

One  question   that  invariably  arises   in  nearly  all   forms  of
observational astronomy: {\it What  is the brightness of that object?}
Therefore many  astronomical display routines  give the user  tools to
answer     this     question     \citep[e.g.~ImExamine    in     IRAF,
  atv.pro\footnote{http://www.physics.uci.edu/$\sim$barth/atv/},
  idp3.pro\footnote{http://mips.as.arizona.edu/MIPS/IDP3}][]{stobie},
and  \igalfit\  is  no   different.   However  a  major  advantage  to
\igalfit\  is the  photometry routines  are integrated  into  an image
display  tool which  combines  the flexibility  of  \dsnine\ with  the
computational power of \texttt{IDL} and the image processing of \sex.

\section{Future Improvements}\label{future}

As with most software, \igalfit\ is a work-in-progress, and there 
are several additions or improvements we would like to include: 

\begin{description}
\item[Asymmetry Parameters:]  In the latest version  of \galfit, there
  are a  bevy of  asymmetry parameters (e.g.~boxiness,  bending modes,
  fourier   components,  truncation   radii)   to  model   azimuthally
  asymmetric structures,  like bars, spiral  arms, or tidal  tails. By
  including  these parameters, the  user can  create far  more complex
  and (perhaps) realistic models.
\item[Additional  Morphological  Programs:]   While  this  project  was
  conceived to provide a user-friendly interface to \galfit, there are
  additional  modeling programs  which can  be included  as  well, for
  example  \texttt{GALPHAT}  \citep{yoon11},  shapelet  decompositions
  \citep[e.g.][]{ref03,massey05},   and  model-independent  estimators
  \citep[e.g.][]{cons03,lotz04,law07}.
\item[Improved Memory  Efficiency and Image Display:]  We mentioned in
  \S~\ref{rungf},  the   rendering  of   large  images  can   be  very
  computationally expensive  and dramatically slow down  even the most
  powerful  workstations.   In  future  versions, we  plan  to  employ
  additional  advanced graphics  capabilities  in IDL  to improve  the
  real-time image display.
\end{description}

\acknowledgments Special thanks to J.~Bosch and M.~Mechtley for advice
and suggestions.   We would also  like to thank our  ``beta testers,''
M.~Rutkowski, S.~Cohen, P.~Thorman, L.~Alcorn, and M.~Jee. Support for
this work  was provided by NASA  through grant numbers  11772 from the
Space Telescope  Science Institute, which  is operated by  AURA, Inc.,
under NASA contract NAS 5-26555.

\begin{appendix}
\section{Example Regions File}\label{regfile}
\igalfit\ will read and write regions files in the same format as \dsnine, 
allowing users to employ existing tasks to define regions.  For completeness, 
we give an example regions file, as written by \igalfit.

\begin{verbatim}
# Region file made by iGalFit on Mon Aug  1 00:31:43 2011
# Filename: psf_f125w.fits
global color=green dashlist=8 3 width=1 font="helvetica 10 normal" select=1 highlite=1 dash=0
 fixed=0 edit=1 move=1 delete=1 include=1 source=1
image
circle(593,586,44.019807) # color=magenta
ellipse(456,543,75,28,338.58853) # color=green
ellipse(679,458,40,20,35.134193) # color=red
ellipse(528,476,40,20,330.01836) # color=blue
box(587.5,547.5,369,313,0) # color=yellow
-box(537.5,605.5,37,59,0) # color=black
-box(713,607,80,40,0) # color=black
\end{verbatim}

\section{Basic \sex\ Catalog}
As  discussed  in \S~\ref{sextractor},  \igalfit\  can  call \sex\  to
identify objects for later use, but can also take a catalog derived by
other  means.  However  to properly  interpret the  columns,  the file
should be in the \texttt{ASCII\_HEAD} format.  here we give an example
of a  catalog which contains the mandatory  fields (additional columns
may be present).
\begin{verbatim}
#   1 NUMBER          Running object number
#   2 ISOAREA_IMAGE   Isophotal area above Analysis threshold         [pixel**2]
#   3 X_IMAGE         Object position along x                         [pixel]
#   4 Y_IMAGE         Object position along y                         [pixel]
#   5 MAG_AUTO        Kron-like elliptical aperture magnitude         [mag]
#   6 MAGERR_AUTO     RMS error for AUTO magnitude                    [mag]
#   7 A_IMAGE         Profile RMS along major axis                    [pixel]
#   8 B_IMAGE         Profile RMS along minor axis                    [pixel]
#   9 THETA_IMAGE     Position angle (CCW/x)                          [deg]
#  10 FLAGS           Extraction flags
         1        17     55.004     12.177  -8.3319   0.0490     1.120     0.930 -21.7   0
         2       118     46.653     48.326 -10.8804   0.0100     2.262     2.078 -17.9   0
         3         7     22.492     47.101  -7.7120   0.0458     0.594     0.569  35.4   0
\end{verbatim}

\end{appendix}

\end{document}